\UseRawInputEncoding
\documentclass{article}
\usepackage{waspaa21,amsmath,graphicx,url,times, tabularx, booktabs, enumitem}
\usepackage{color}

\def \am [#1]{\textcolor{red}{AM: #1}}
\def \rev [#1]{\textcolor{blue}{#1}}
\def \im [#1]{\textcolor{ForestGreen}{IM: #1}}

\title{CROWDSOURCING STRONG LABELS FOR SOUND EVENT DETECTION}
\name{Irene Mart\'{\i}n-Morat\'o, Manu Harju, Annamaria Mesaros\thanks{This paper has received funding from Academy of Finland grant 332063 ``Teaching machines to listen".}}

\address{Computing Sciences,
    Tampere University\\
    Korkeakoulunkatu 7, 33720 Tampere, FINLAND\\
    \{irene.martinmorato, manu.harju, annamaria.mesaros\}@tuni.fi
}

\begin{document}

\ninept
\maketitle

\begin{sloppy}

\begin{abstract}

Strong labels are a necessity for evaluation of sound event detection methods, but often scarcely available due to the high resources required by the  annotation task. We present a method for estimating strong labels using crowdsourced weak labels, through a process that divides the annotation task into simple unit tasks. Based on estimations of annotators' competence, aggregation and processing of the weak labels results in a set of objective strong labels. The experiment uses synthetic audio in order to verify the quality of the resulting annotations through comparison with ground truth. The proposed method produces labels with high precision, though not all event instances are recalled. Detection metrics comparing the produced annotations with the ground truth show 80\% F-score in 1~s segments, and up to 89.5\% intersection-based F1-score calculated according to the polyphonic sound detection score metrics.

\end{abstract}


\section{Introduction}
\label{sec:intro}

The research on sound event detection is currently dominated by methods that learn acoustic models from weakly-labeled data \cite{Kong2020, Turpault2019_DCASE}, in which only presence of sound events is indicated, without explicit temporal information. However, the task of sound event detection is defined as recognizing and temporally locating sound instances within a recording \cite{mesaros2018datasets}, which creates a mismatch between the training and the requirements imposed on the system. The main cause of this situation is the lack of suitable datasets containing strong labels, that indicate both textual labels and temporal boundaries of events. 

While training of acoustic models can be achieved with advanced learning methods using weakly-labeled real-life recordings or strongly-labeled synthetic audio mixtures, the evaluation of such methods still requires strongly-labeled data. Typically, a small amount of data is manually annotated for evaluation. In recent data evaluation challenges, the evaluation data consisted of short recordings, of length 10 seconds \cite{Turpault2019_DCASE},  while earlier ones used recordings of 3-5 minutes \cite{Mesaros2019_TASLP}. Such annotation efforts were concentrated to individual research groups, resulting in datasets of small size. The most recent work introduces strong labels for a part of AudioSet \cite{Gemmeke2017}, providing 67k manually annotated clips. The length of these clips is 10 s, and they contain on average 3.5 labels \cite{Hershey2021_ICASSP}. 

Manual annotation of sound events is subjective in many ways, from the textual labels selected for the sound \cite{guastavino2018}, to the placing of temporal boundaries for the event instances \cite{mesaros2018datasets}.
Ideally, an objective reference annotation is based on multiple annotators or curation, but curation of strong labels is difficult. For example in \cite{Hershey2021_ICASSP}, a first-pass labeling was reviewed by a different annotator who could modify the labels, but even with 5 stages this process rarely converged to consensus. On the other hand, multiple, independent annotators should be followed by a method for aggregating the information. In image analysis, aggregation is usually done as intersection of segments \cite{Kauppi2009} or by maximizing agreement between annotators \cite{Kamarainen2012}. For audio, it is rare to have data with multiple annotators; in particular, strong labeling is impractical for multi-annotator solutions because of the difficulty of the task.

An attractive solution for annotating large amounts of data, increasingly used for audio annotation, is crowdsourcing \cite{Cartwright2019, fonseca2019, humphrey2018openmic}. Crowdsourcing is suitable for simple annotation tasks like classification, with a number of services offering ready-made solutions for it, but these are not suitable in their current form for strong labeling. 

In this paper, we propose a method for creating strong labels based on weak labels of overlapping segments. Weak labeling allows use of crowdsourcing, facilitating annotation of high volume of audio files in a fast way.  
The novelty of this work is threefold:

\noindent 1. We introduce a novel, proof of concept, method for crowdsourcing strong labels, by estimating the strong labels based on audio tags (weak labels); 2. We evaluate the correctness of the collected annotation with respect to the ground truth, by using synthetically generated audio mixtures for which the reference annotation is generated at the same time with the audio mixtures; 3. We investigate the effect of reference annotations on the evaluated performance of sound event detection systems by using the ground truth annotations in training, but evaluating against the manually created reference annotation. 

We show that it is possible to obtain reasonable strong labels by crowdsourcing segment-level tags and further processing them. 
The resolution of the estimated labels is determined by the degree of overlap of consecutive segments.
We use annotator competence estimation \cite{hovy2013} to eliminate the poor quality answers from the collected data before further processing. 
We also show that the mismatch between the synthetic ground truth and the manually created annotation produces a significant drop in measured performance, even though it does not affect system functionality.
All data produced and collected in this study is publicly available.\footnote{MAESTRO Synthetic, 10.5281/zenodo.5126478} 
The paper is organized as follows:
Section \ref{sec:ann_method} presents the proposed annotation procedure and data processing to estimate the strong labels. Section \ref{sec:experiments} presents the experimental setup, dataset, annotation process, and analysis of the collected annotations. Section \ref{sec:sed_sys} shows the use of crowdsourced annotations in evaluation of a sound event detection system. Finally, Section \ref{sec:concl} presents conclusions and future work.

\section{Annotation method and processing}
\label{sec:ann_method}

\subsection{Annotation method}

In order to elicit a consistent behavior from annotators and produce a consistent output, an annotation task should require a single, simple decision \cite{mesaros2018datasets}. Annotating audio with strong labels by requiring from the annotator textual labels for sounds, along with onset and offset for each sound instance, is the exact opposite.
We propose a procedure that divides the strong annotation task into unit tasks that involve simple decision-making: to indicate presence of sounds from a pre-selected list labels in pre-segmented audio \cite{mesaros2018datasets}, practically weakly labeling individual segments. In addition to simplifying the work of the annotator, this approach makes the annotation task suitable for crowdsourcing. The audio files are segmented into short, overlapping segments, which are then annotated with weak labels by indicating binary activity of sound events within the entire segment. The list of target sound classes is selected in advance and presented to the annotator.
The proposed method, illustrated in Fig. \ref{fig:method}, uses a sliding ``annotation window" over the length of the audio file, with a high rate of overlap between consecutive segments. The temporal activity of sounds within the original long file can then be estimated based on the tags of the individual segments, by reconstructing the temporal sequence of these segments into an aggregated representation that counts activity indicators at each time step. If all annotations are correct, event boundaries correspond to the boundaries of the maximum-valued region in the count-based activity indicators.  

\begin{figure}
    \centering
    \includegraphics[width=0.8\columnwidth]{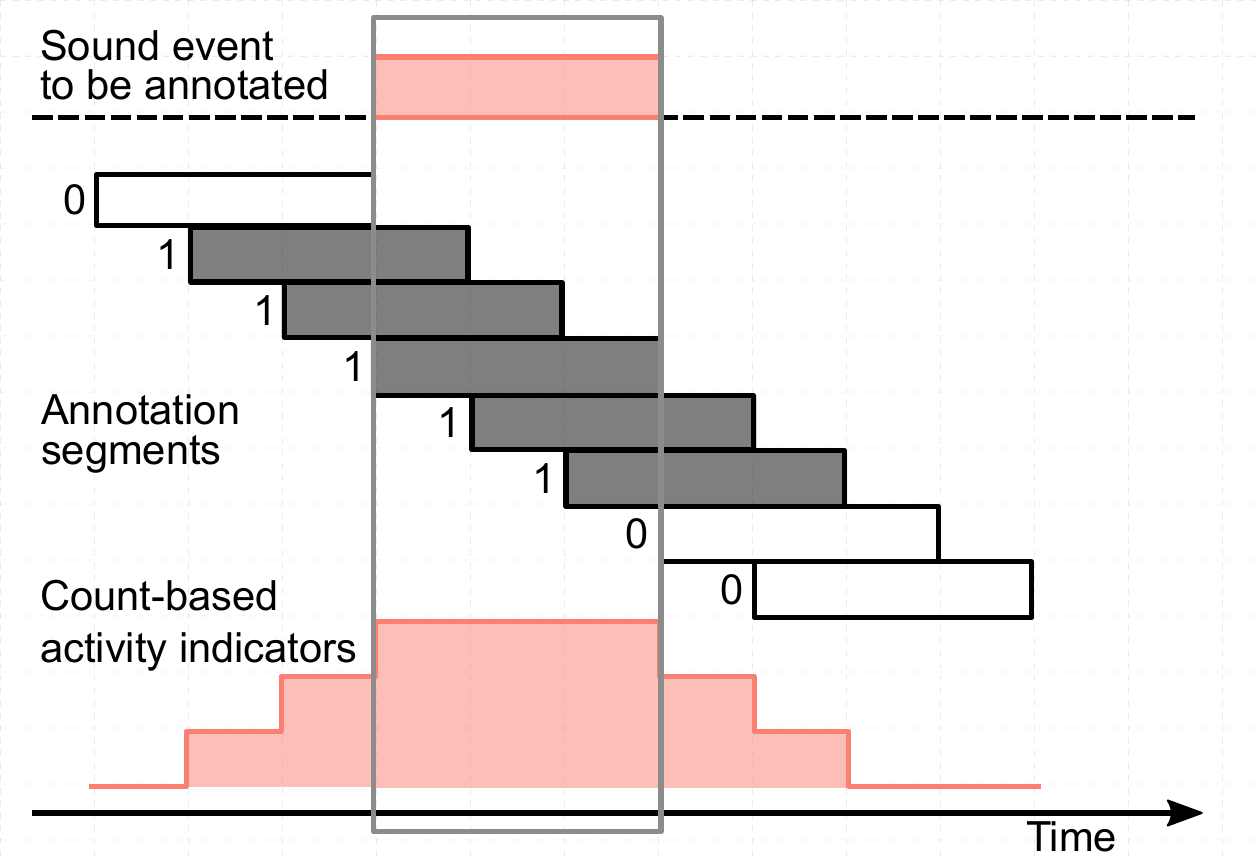}
    \caption{Estimating event activity from overlapping weakly-labeled segments.}
    \label{fig:method}
   \end{figure}

In this work, we choose a segment length of 10 seconds. We rely for this choice on studies that show accurate recognition of sound sources to be at most 6.8 seconds for a list of 42 different sounds \cite{Ballas1993}. With a hop of one second between the segments, the temporal reconstruction of the events activity will have a one second resolution, similar to the diffuse labels created in \cite{Hershey2021_ICASSP}.
Following the procedure from \cite{Cartwright2019}, which was also used in \cite{martinmorato2021}, the task is presented as a single-pass multi-label annotation, in which the presence of a sound is explicitly indicated, and the absence is implicit by the label not being selected. 

\subsection{Annotator competence and ground truth estimation}

Multiple annotator opinions are typically aggregated using majority vote in order to estimate reference labels for the data \cite{Cartwright2019, humphrey2018openmic}. In this work we use MACE - Multi-Annotator Competence Estimation \cite{hovy2013} to identify the trustworthy annotators and to predict the labels. MACE models the behavior of the annotators in order to estimate the competence of the annotators and the true labels of the data. Annotator behavior is modeled using a binary variable drawn from a Bernoulli distribution, encompassing the annotator trustworthiness and the spamming behavior. The true labels and the spamming indicators are estimated based on the observed annotations, using expectation maximization. For complete details on the model assumptions and the method, we refer the reader to \cite{hovy2013}. In the estimation of the true labels, the annotators' opinions are weighted based on their competence, in contrast to majority voting where each annotator's opinion has the same weight. 
The method has been extended for the audio tagging scenario in our previous work, by considering each multi-labeled item as a set of binary \textit{yes/no} labels per item \cite{martinmorato2021}. Each (item, label) pair is used as a separate annotator opinion, for which MACE estimates the true label. This approach models the single-pass multi-label annotation as a multiple-pass binary annotation \cite{Cartwright2019}. The method was shown to recognize well the spamming behavior of annotators in audio tagging \cite{martinmorato2021}, providing a satisfactory level of inter-annotator agreement when the least competent annotators' opinions are removed from the data pool.

\section{Experimental Setup and results}
\label{sec:experiments}

\subsection{Dataset and annotation procedure}
\label{sec:dataset}

Audio files are generated using Scaper \cite{salamon2017}, with small changes to the synthesis procedure.
A soundscape is generated by placing events iteratively at random intervals until the desired maximum polyphony of 2 is obtained. Intervals between two consecutive events are selected at random, but limited to 2-10 seconds. Event classes and event instances are chosen uniformly, and mixed with a signal-to-noise ratio (SNR) randomly selected between 0 and 20 dB over a Brownian noise background. Having two overlapping events from the same class is avoided.
Foreground events are extracted from the UrbanSound dataset \cite{salamon2014}. The dataset includes classes: car\_horn, children\_playing, dog\_bark, engine\_idling, siren, and street\_music, with children\_playing renamed to children\_voices for the annotation task, and files shorter than one second or longer than 60 seconds discarded.  The classes were selected to mimic the street scenes annotated in \cite{martinmorato2021}.
Dataset statistics are presented in Table \ref{tab:events}. 

\begin{table}[]
\centering
    \begin{tabular}{l|c}
    \toprule
    event class & instances \\
    \midrule
    car horn & 109 \\
    children voices & 236 \\
    dog bark & 343 \\
    engine idling & 564 \\
    siren & 256 \\
    street music & 89 \\
    \bottomrule
    \end{tabular}
    \caption{Number of instances of each event class in the data}
    \label{tab:events}
    \vspace{-10pt}
\end{table}

The dataset consists of 20 generated soundscapes, each having a length of 3 minutes. The resulting files are cut into 10 second segments with 1 second offsets, resulting in 171 segments from a single 3 minute soundscape, and a total number of 3420 10-second clips to be annotated.
Each individual 10~s segment was considered as an independent annotation task, provided on Amazon Mechanical Turk as one HIT (Human Intelligence Task). In order to prevent the same worker annotating overlapping segments, the data was split into batches containing segments located at least 15 seconds apart in the original audio. The batches were then launched one at a time, and workers that already performed at least 50 hits in previous batch(es) were disqualified from working on the task. A payment of \$0.10 was offered per HIT. Worker qualification was requested as at least 1000 completed HITs with average approval rating of at least 85\%.
One HIT consisted of listening to the provided audio excerpt and indicating which sounds are present in it, from the given list of six classes or ``none of the above". The number of playbacks allowed was not limited. No visualization (e.g. spectrogram) was provided. Workers were instructed to complete the task using headphones, and in a quiet environment. Before the job, they were also provided short descriptions for every class, and four example sounds that contained events from all classes.
The complete data annotation task was performed by 680 workers, with each 10~s segment being annotated by 5 workers. All jobs were accepted, in order to study the annotator behavior. 

\subsection{Analysis of annotation outcome}
\label{sec:analysis}

The correctness of the collected audio tags with respect to the generated reference labels was evaluated by considering the annotated segments individually. When the multiple opinions per segment are aggregated through majority vote, the comparison of the resulting tags with the ground truth achieves 68\% F-score, with 98\% precision, 52\% recall. Using MACE to predict the true labels provides 86\% F-score, with 97\% precision and 77\% recall, while union results in 78\% F-score, with 70\% precision and 89\% recall.
The relatively small recall indicates that many sounds are not annotated, possibly not being perceived in the audio mixture. With the majority vote, only slightly over half of the tags are found, while MACE raises the number of the tags correctly recalled to over three quarters. On the other hand, the very high precision indicates that the sounds which are labeled are labeled correctly. To keep this study focused on the annotation method itself, we do not analyze here the influence of SNR on recall. We note, however, that similar annotator behavior was observed as polyphony increases for annotators that selected onset and offset of sound instances \cite{cartwright2017seeing}.

\begin{figure}
    \centering
    \includegraphics[width=0.85\columnwidth]{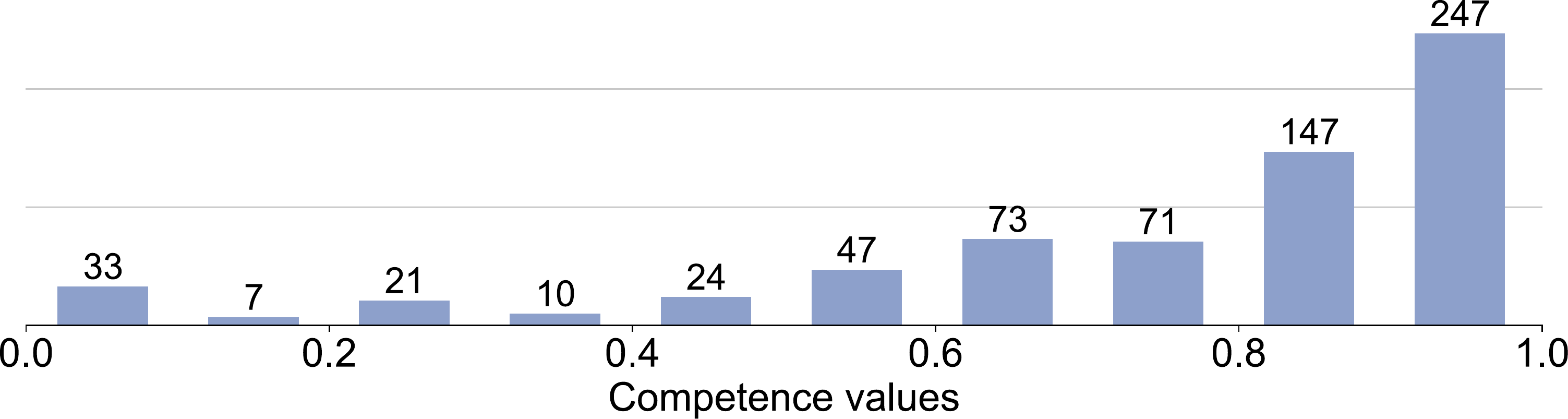}
    \caption{Annotator competence estimated using MACE. }
    \label{fig:annotator-competence}
\vspace{-8pt}
\end{figure}

The annotator competence analysis performed by MACE, presented in Fig.~\ref{fig:annotator-competence}, reveals that 33 annotators have answered randomly, while 247 have a competence over 0.9. Inter-annotator agreement, measured using Krippendorff's alpha is 0.57. 
Eliminating annotators with low competence increases the agreement, e.g. a competence threshold of 0.6 increases $\alpha$ to 0.72, while a threshold of 0.8 increases $\alpha$ to 0.80. 
For further experiments, we use a competence threshold of 0.6, which keeps approximately 80\% of annotators (538 of 680 annotators).

\subsection{Estimation of strong labels}
\label{sec:est_strong_lab}

The temporal activity patterns of sound events is constructed as explained in Section \ref{sec:ann_method}, by stacking the annotated segments in their original order. 
In the basic setup, 5 opinions per 10~s segment result in 50 opinions for each 1~s of the estimated temporal activity (except the first and last 10~s of the original files).
We estimate the temporal activity in three ways:
\begin{enumerate}[nolistsep]
    \item using data from all annotators, as explained above, with 50 opinions per 1~s segment.
    \item using only annotators with competence higher than 0.6 (obtained using MACE). In this case, due to eliminating some of the annotations, the number of opinions per 1~s segment varies, being 37 on average 
    \item using the tags estimated using MACE as explained in Section \ref{sec:ann_method}. In this case there is one opinion per 10~s segment (the MACE output), resulting in 10 opinions per 1~s. 
\end{enumerate}

\noindent 
The resulting representation is then binarized using a threshold applied in each 1~s segment. Instead of the theoretical maximum value M for each 1~s, we use a threshold of 80\%, to accommodate possible incorrect answers from the annotators. This reflects the proportion of annotators with an estimated competence over 0.6, as presented in Section \ref{sec:analysis}. A sound event is therefore considered active in a 1~s segment if at least 80\% of opinions available for that segment considered it active. 
Fig.~\ref{fig:label-estimation} presents an example of this process, along with the ground truth for comparison.
The quality of the resulting strong labels is evaluated by calculating detection metrics between them and the ground truth. For this, we calculate ER and F1 in 1~s segments (ER\_1s, F1\_1s), following the sound event evaluation procedure from DCASE Challenge \cite{Mesaros2019_TASLP}, and the intersection-based F-score as defined in \cite{Bilen_2020}. For the latter, we use two scenarios, as defined in DCASE 2021 Challenge Task 4\footnote{http://dcase.community/challenge2021/task-sound-event-detection-and-separation-in-domestic-environments\#evaluation}, with different criteria for the detection tolerance criterion (DTC) and ground truth intersection criterion (GTC): DTC=GTC=0.7 (F1\_dtc=0.7) and DTC=GTC=0.1 (F1\_dtc=0.1). For details on the parameters and their effect, we refer the reader to \cite{Bilen_2020}. 
The results are presented in Table \ref{tab:analysis}.  

\begin{figure}
    \centering
    \includegraphics[width=0.9\columnwidth]{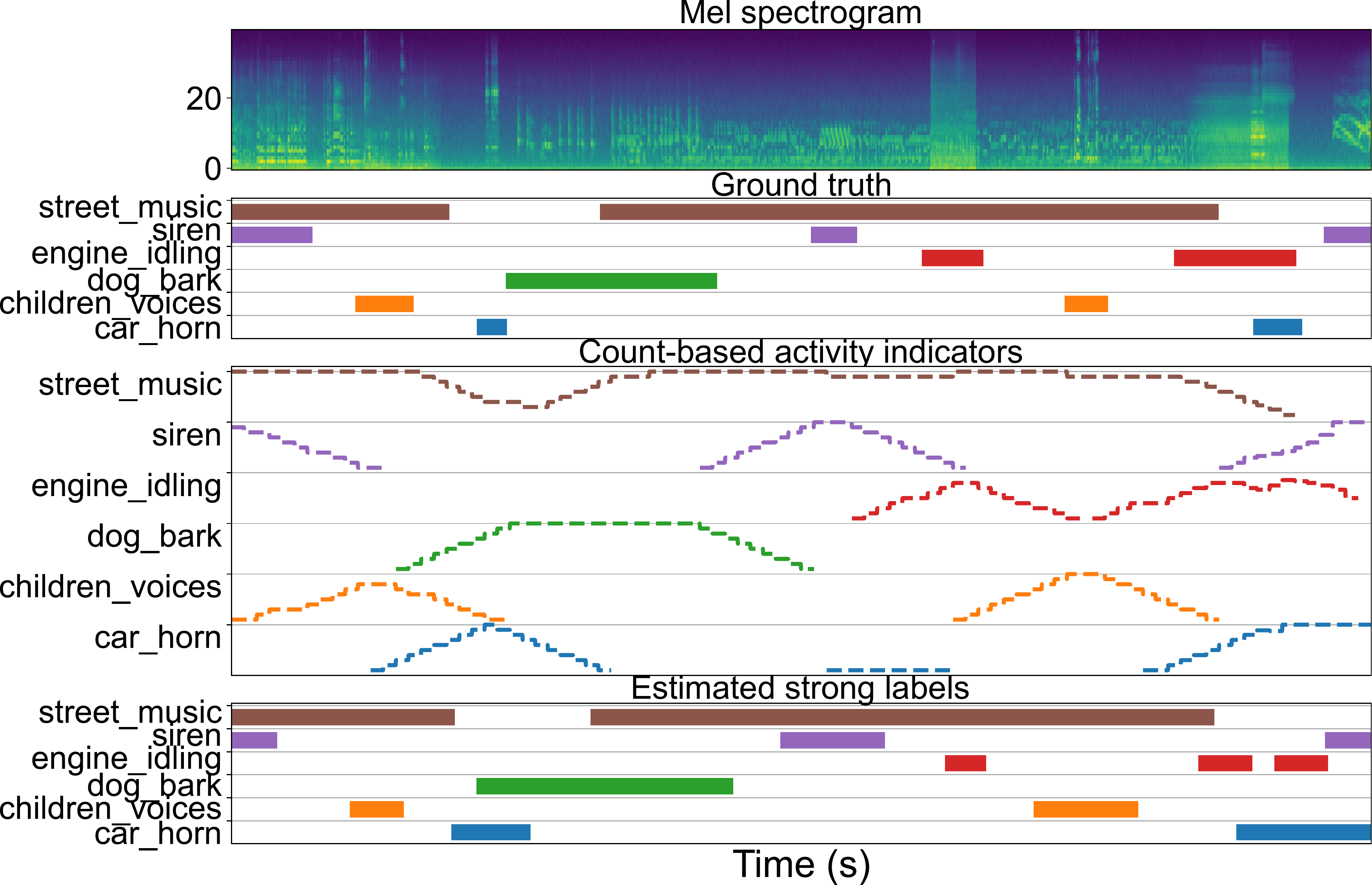}
    \caption{Estimation of strong labels based on the weak labels of consecutive, overlapping, audio segments. }
    \label{fig:label-estimation}
\end{figure}

\begin{table*}[]
    \centering
    \begin{tabular}{l|c |c c c|| c| c c||c c}
    \toprule
     labels based on    & ER$_{1s}$ & S & D & I & F1$_{1s}$ & P & R & F1$_{dtc=0.7}$ & F1$_{dtc=0.1}$\\
     \midrule
    all annotators      & 0.55  & 0.01  & 0.51  & 0.03  & 62.6 & 91.2  & 47.7 & 39.0\% & 67.4\%\\ 
    competence $>$ 0.6  & 0.44  & 0.02  & 0.36  & 0.05  & 72.6 & 88.9  & 61.3 & 44.0\% & 81.3\% \\
    MACE                & 0.36  & 0.02  & 0.19  & 0.14  & 80.1 & 82.3  & 77.9 & 41.2\% & 89.5\% \\
    \bottomrule
    \end{tabular}
    \caption{Sound event detection scores between the estimated strong labels and the ground truth.}
    \label{tab:analysis}
    \vspace{-8pt}
\end{table*}

A comparison of the estimated and the ground truth labels is presented in Fig.~\ref{fig:estimation-example}. This example shows that the proposed method has difficulty in estimating correctly the temporal activity for the short sound events. Even though most of the sounds in the example are identified, the high mismatch between the temporal boundaries of the short sounds will increase segment-based error rate and decrease F-score (as false positives or insufficient intersection).
A more lenient intersection criterion (PSDS with DTC=0.1) results in an F-score of almost 90\% for the best case. 

\subsection{Discussion}

A close analysis of the detection scores reveals that when relying on the majority vote among all annotators, the error rate is composed mostly of deletions, with only a small proportion of insertions and substitutions. This was expected based on the results from section \ref{sec:analysis}, which indicated that many sounds were not annotated (recall 52\%). 
MACE introduces many labels compared to the other methods, thus reducing deletions, but creates insertions because not all these labels are correct. 
The same trend can be seen in the dynamics shift of precision and recall: while the labels estimated based on all annotators have a high precision of 91.2\%, but only 47.7\% recall, MACE obtains a recall of 77.9\% at the cost of reducing precision to 82.3\%. 
Since tags produced by MACE for the 10~s segments had 86\% F-score, with about one quarter of tags missing (recall 77\%), a detection F-score of 89.5\% (80.1\% segment-based) between the estimated strong labels and ground truth ones means a very good match, provided that the missing tags may correspond to sound instances that were not perceived by the annotators.

\begin{figure}
    \centering
    \includegraphics[width=\columnwidth]{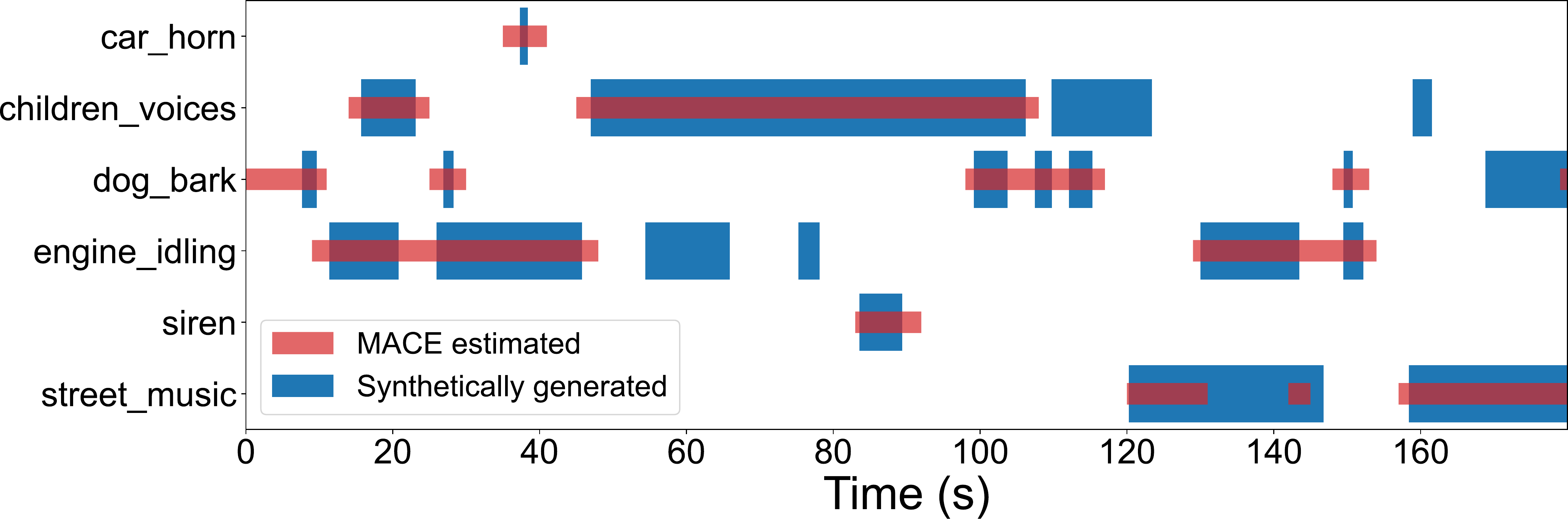}
    \caption{Visual comparison of generated and estimated labels}
    \label{fig:estimation-example}
    \vspace{-10pt}
\end{figure}

\section{Sound event detection}
\label{sec:sed_sys}

As an additional experiment, we test how the mismatch between the estimated and synthetically generated strong labels affects SED evaluation. We use the system ranked best in the sound event detection task of DCASE 2017, where a similar amount of data was available. The system is a simple CRNN composed of 3 convolution blocks, each followed by batch normalization and max-pooling layers. The final two layers are composed of bi-directional gated recurrent units (GRU), in order to learn the temporal activity patterns. For more details of the model, we refer the reader to \cite{Adavanne2017}. 
The system is trained using the audio data and the ground truth labels generated using Scaper. In order to use as much training data as possible, the train/test procedure follows a leave-one-out setup, in which one file is kept for testing, in turn, and the other 19 files are used for training and validation (18 for training, one for validation). The system output is evaluated against three different types of reference annotations: 
\begin{enumerate}[nolistsep]
    \item Generated strong labels (ground truth)
    \item Strong labels estimated using annotators with a competence higher than 0.6 (case 2 from Sec. \ref{sec:est_strong_lab})
    \item Strong labels estimated using MACE (case 3 from Sec. \ref{sec:est_strong_lab})
\end{enumerate}

\begin{table}[]
    \centering
    \begin{tabular}{l|c c | c c}
    \toprule
eval. reference & ER${_1s}$ & F1${_1s}$ & F1$_{dtc=0.7}$ & F1$_{dtc=0.1}$ \\ 
\midrule
GT  & 0.49 & 64.8\% & 26.4\% & 39.6\% \\
estim.comp.$>$0.6 & 0.78 & 49.1\% & 12.5\%  & 31.2\% \\
estim. MACE  & 0.65 & 52.2\% & 13.0\% & 31.1\% \\
\midrule
\midrule
train\&eval MACE & 0.58 &55.9\%&  16.4\%& 40.3\% \\
\bottomrule
    \end{tabular}
    \caption{Evaluation against different sets of strong labels}
    \label{tab:sed}
    \vspace{-6pt}
\end{table}

\noindent The results, evaluated using detection metrics, are presented in Table \ref{tab:sed}.
We consider as a baseline performance the system trained and evaluated using the generated labels. Its error rate and F-score align well with the performance reported on other synthetic data, e.g. UrbanSED (F1$_{1s}$ approximately 60\%) \cite{salamon2017} and DCASE 2016 synthetic audio task (top systems had ER$_{1s}$ 0.33-0.40 and F1$_{1s}$ 78-80\%) \cite{Mesaros2018_TASLP}. When evaluated against the human-produced labels, the drop in measured performance is significant, even though what we evaluate is the exact same system output. 
If the same system is trained and evaluated using the estimated strong labels based on MACE, the measured performance is closer to the baseline performance (last row in Table \ref{tab:sed}). However, this system is trained and tested on approximately half the sound instances, as indicated by the low recall of the annotation process.
These results show once more that the quality of annotations is a limiting factor not only in the training stage, but also for performance evaluation. 
The presented experiment is a typical situation, training SED systems on synthetic audio with correct and complete strong labels for training, and testing it on real-life recorded data. In addition to the presented effect of incomplete labels, testing on real recordings will introduce errors due to the mismatch in acoustic data. The presented system is a rather simple one, not considered state-of-the-art, therefore the effects of a weak system and the incomplete annotation are combined in the evaluated performance. However, we expect a similar effect of the annotations on more powerful systems too.

\section{Conclusions}
\label{sec:concl}

As sound event detection applications are moving towards systems applicable in real-life, a limiting factor of the development is the data annotation process. Even though training of systems can be achieved without strongly-labeled data, manually annotated data is necessary for evaluating the system behavior on real recordings corresponding to the user scenario. To alleviate the burden and subjectivity of manual annotation, we presented a method that can produce strong labels through crowdsourcing. Based on annotator competence estimation, a good, though incomplete, set of labels was produced. The resulting aggregated annotation is objective, being composed of multiple opinions. In future work, we will investigate further optimization of MACE for the case of strong labels, and investigate methods for producing the minimum amount of labels necessary for a reliable estimation, to reduce the redundancy of annotations where possible.

\newpage

\bibliographystyle{IEEEtran}
\bibliography{refs21}

\end{sloppy}
\end{document}